\documentclass{nature}
\usepackage{setspace}
\usepackage{amsfonts}
\usepackage{amssymb}
\usepackage{amsmath}
\usepackage{float}
\usepackage{placeins}
\usepackage{graphicx}
\title{Data Bias in Human Mobility is a Universal Phenomenon but is Highly Location-specific}

\author{Katinka den Nijs$^{1}$, Elisa Omodei$^{2}$, and Vedran Sekara$^{1,3*}$}

\begin{document}

\maketitle

\begin{affiliations}
 \item Networks, Data, and Society (NERDS) group, IT University of Copenhagen, Copenhagen, Denmark
 \item Department of Network and Data Science, Central European University, Vienna, Austria
 \item Pioneer centre for AI (P1), Copenhagen, Denmark
 \item[]$^{*}$To whom correspondence should be addressed vsek@itu.dk
\end{affiliations}

\begin{abstract}
Large-scale human mobility datasets play increasingly critical roles in many algorithmic systems, business processes and policy decisions.
Unfortunately there has been little focus on understanding bias and other fundamental shortcomings of the datasets and how they impact downstream analyses and prediction tasks.
In this work, we study `data production', quantifying not only whether individuals are represented in big digital datasets, but also how they are represented in terms of how much data they produce.
We study GPS mobility data collected from anonymized smartphones for ten major US cities and find that data points can be more unequally distributed between users than wealth.
We build models to predict the number of data points we can expect to be produced by the composition of demographic groups living in census tracts, and find strong effects of wealth, ethnicity, and education on data production.
While we find that bias is a universal phenomenon, occurring in all cities, we further find that each city suffers from its own manifestation of it, and that location-specific models are required to model bias for each city.
This work raises serious questions about general approaches to debias human mobility data and urges further research.
\end{abstract}

% ------------------- INTRODUCTION -------------------

\section*{Introduction}
Our capacities to collect, store, and analyze vast amounts of human mobility data have greatly increased in the past decades~\cite{lazer2009social}.
Today, human mobility data is used by researchers, governments, and business for a wealth of purposes including: urban planning~\cite{wang2012understanding}, estimating and addressing poverty~\cite{blumenstock2018estimating}, dynamic population mapping~\cite{deville2014dynamic}, tourism estimation~\cite{demunter2017tourism}, quantifying migrations after sudden onset emergencies~\cite{acosta2020quantifying}, and epidemic forecasting of various diseases such as COVID-19~\cite{bonaccorsi2020economic,schlosser2020covid}, Ebola~\cite{merler2015spatiotemporal}, and Malaria~\cite{wesolowski2012quantifying}.
Human mobility datasets are collected through various technological means. In our increasingly digitalized world, large scale mobility data is often passively collected as a by-product of digital technologies used for billing, service, or marketing purposes. This includes: call and extended detail records collected through regular mobile phone usage, GPS traces collected via apps installed on smartphones, check-ins on online social media, and smart travel card data.
These data collection technologies naturally introduce some forms of bias in the data~\cite{salganik2019bit}.
In fact, it is widely recognized that access to, and usage of these big data technologies varies across populations and that not all demographics are equally represented in these datasets~\cite{sekara2019mobile,sapiezynski2020fallibility,li2024understanding}.
For instance, in the US around 91 out of 100 adults own a smartphone, with younger, wealthier, and college educated groups being more likely to own one~\cite{sheet2024pew}.
However, even when marginalized populations own digital devices (e.g. smartphones) individuals might limit their behavior, resulting in them producing fewer data points and generating data of lower utility~\cite{schlosser2021biases}. 
As such, it remains an open question how truthfully these large digital datasets represent the actual travel behavior of the general population.

Here, we study the representativeness of high-resolution human mobility datasets.
Our focus is not only on whether different demographic groups are included in these datasets, but also how they are included, i.e. how much data do different groups produce and contribute to mobility datasets.
We call this \textit{data production bias}, and we focus on this type of representativeness for multiple reasons.
First, the number of data points an individual contributes to a dataset has direct influence on the quality of mobility traces, i.e. the more data points people produce the more complete travel networks can be inferred.
Second, for data cleaning it is common practice to filter away individuals with few data points, but the question is which individuals are disregarded, and are specific demographics removed more often than others?
Third, to be able to fix biases in mobility data, we need a better understanding of which groups are in these datasets, and how they are represented.
To understand data representativeness, we focus on the aggregated amount of data produced within a specified time frame.
We use anonymized mobility traces (on GPS resolution) collected using smartphones and provided by a mobility data provider.
The provider deliver anonymized records of GPS locations from users who have opted-in to provide access to their location data, while remaining compliant under the General Data Protection Regulation and the California Consumer Privacy Act.
We focus on mobility data from the ten largest cities by population in the United States during one representative month, April 2019, chosen prior to COVID-19 to avoid any effects on mobility patterns induced by non-pharmaceutical measures aimed at combating the pandemic, and also potential post-recover effects.

\section*{Results}
\begin{figure}[!htbp]
\includegraphics[width=1\linewidth]{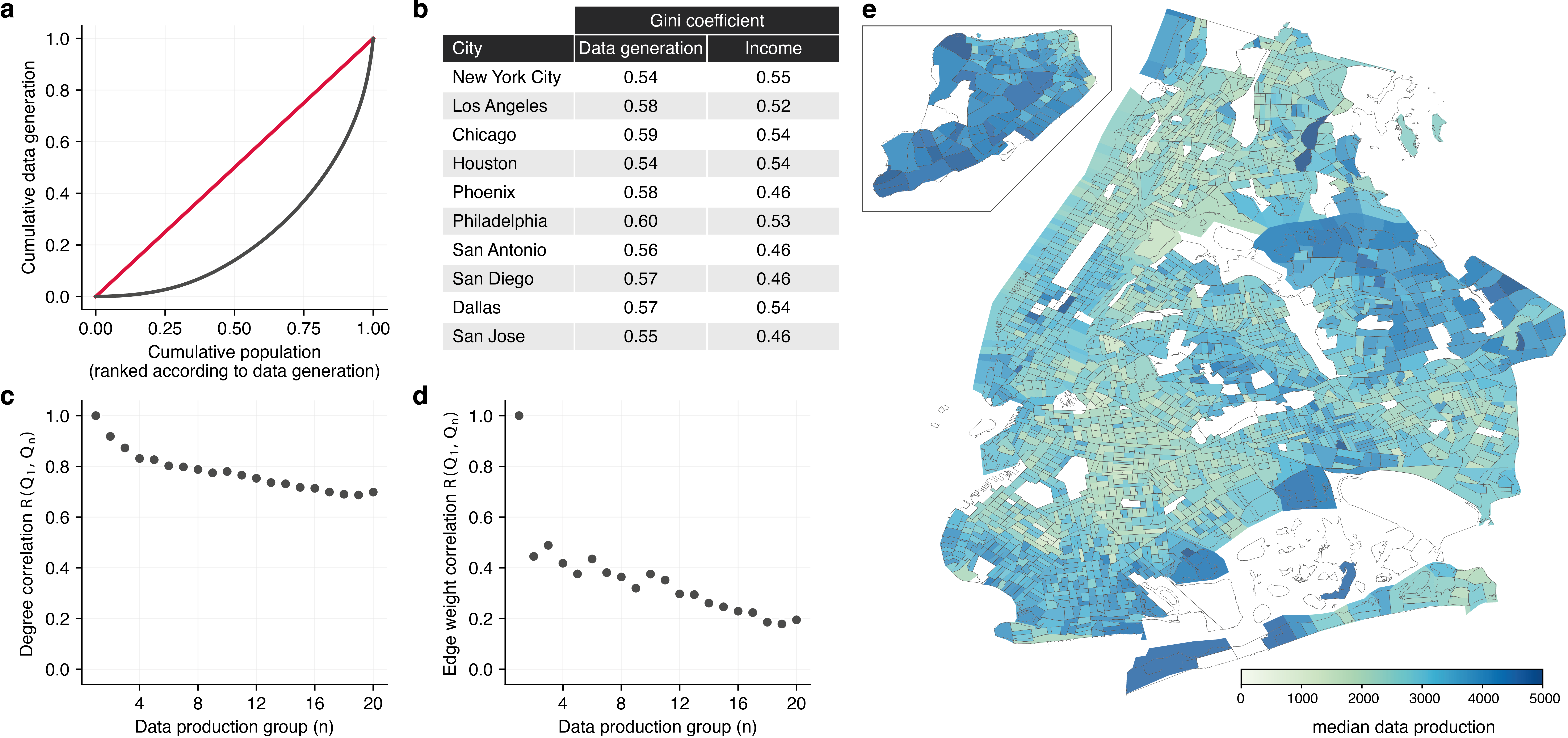}
\caption{\textbf{Observing biases in mobility data.} \textbf{a}, Lorenz curve for the cumulative data production distribution for New York City. The black line denotes the observed distribution, with an associated Gini index of 0.54, while the red line denotes the line of equality. The closer the black line is to the red, the more equally data is produced by individuals. \textbf{b}, Gini indexes for the data production distributions compared to Gini indexes for income distributions in the 10 most populous US cities in 2020. Gini indexes for income are from the American Community Survey (ACS) for 2019~\cite{acs2019}. \textbf{c}, Pearson correlations between degree distributions from the twenty mobility networks, constructed by aggregating trips from individuals in each group. Data shows the travel behavior for New York City. Group one indicates the top data producers, while group twenty contains the lowest producing individuals. Here we compare all networks to group one, containing the highest data producing individuals. \textbf{d}, Edge weight correlations (Pearson) between mobility networks from the twenty groups compared to the network from group one. \textbf{e}, Data production map of New York City showing the median number of datapoint produced by inhabitants with home-locations in each census tract during April 2019. Inset shows Staten Island.}
\label{fig:fig1}
\end{figure}

It is well-documented that wealth, income~\cite{chancel2022world}, access to education and opportunity~\cite{olsen2022unequal}, and access to health care~\cite{fleurbaey2009unfair} are all unequally distributed.
Counting the number of data points individuals produce in our human mobility dataset we find a similar distribution (Fig.~\ref{fig:fig1}a).
For New York City a majority of individuals (80\%) produce less than half of all data points ($\sim$40\% of the data), while a minority (20\%) produce approx. 60\%. 
Put differently, mobility datasets contain travel patterns for a multitude of people, but the behavior of a small minority dominates these datasets.
This undoubtedly skews the data.
In fact, the Gini index for the data production distribution is 0.54, which is remarkably similar to the income distribution for the city (Gini-index 0.55~\cite{acs2019}).
As such, data points in large scale human mobility datasets are as unequally distributed as wealth.
The data production distribution for New York City is not an outlier (Fig.~\ref{fig:fig1}b).
Other large cities in the US have comparably high Gini indices for their data distributions (see SI Section S1).
Further, we observe that for most of these cities, mobility data is even more unequally distributed than income, e.g. Philadelphia has a Gini index of 0.6 for data production, while having a Gini index of 0.53 for income. 

Unequal data production will leave marks in the associated collective mobility networks, which are derived from individual traces.
To investigate this we divide individuals up into 20 equally sized groups ($Q_1$ to $Q_{20}$) according to their data production levels (such that each groups contains 5\% of individuals in our dataset), and generate collective mobility networks for each of these groups by aggregating individuals trips.
Trip start and end locations are inferred at census tract level, such that nodes in the networks are individual tracts, and edges represent the number of trips individuals have made between pairs of census tracts.
We reduce noise by removing edges which have a weight less than two (i.e. collectively individuals need to have made at least two trips between a pair of census tracts for the edge to be included in the analysis), and by extracting the noise-corrected backbone~\cite{coscia2017network}.
To understand how $Q$-networks relate to each other we calculate the degree correlation between the $Q_1$ network and the rest (Fig.~\ref{fig:fig1}c).
The figure shows that census tracts which have a high degree in the mobility network generated by group one ($Q_1$, highest data producing individuals) tend to also have a high degree in other networks (and vice versa for low degree tracts). 
This indicates that structurally these networks resemble each other.
However, when looking at edges weights to understand how strongly census tracts are connected, we find there are only weak correlations between networks (Fig.~\ref{fig:fig1}d). 
This demonstrates that although networks have similar topology, the underlying strength of the paths are remarkably different.
For instance, groups $Q_1$ and $Q_2$ have remarkably few trips in Manhattan compared to $Q_{19}$ and $Q_{20}$, while having considerably more trips in Staten Island (see SI Fig. S2).

Differences in networks can be caused by different factors.
One component of this can be tied with geographic or demographic factors; for example, high data producing individuals can live geographically separated from other groups, be wealthier, or it can be combinations of these.
To uncover relationships between data production and demographic factors we link mobility patterns to census data.
Unfortunately, the data provider does not have any demographic information for individual users.
As such, we link all individuals in our study to census tracts through inferred home locations (see Methods), and each individual is linked to one `home census tract'.
For census tracts there exists rich demographic information from the United States Census Bureau, describing characteristics such as: poverty, sex, age, fraction of university educated individuals, ethnicity, etc.
As we reduce the granularity of our data from user to census tract level we move away from looking at individual level data to studying collective behavior. 
For each census tract we quantify data production as the amount of data produced by the median individual living there (see Methods).
This effectively reduces noise as `outlier individuals' have little effect on the median.
Fig.~1e shows the geospatial distribution of data production rates for New York City, where individuals only contribute to their home census tract, illustrating that median data production levels are not random, rather they vary spatially throughout the city.
For example, large parts of Brooklyn have low median data production levels, while the inhabitants from Staten Island have high levels.
This calls for a closer examination of the relationship between data production and census tract demographic composition.

\subsubsection*{Uncovering factors that contribute to data production bias}
\begin{figure}[!htbp]
\centerline{\includegraphics[width=0.7\linewidth]{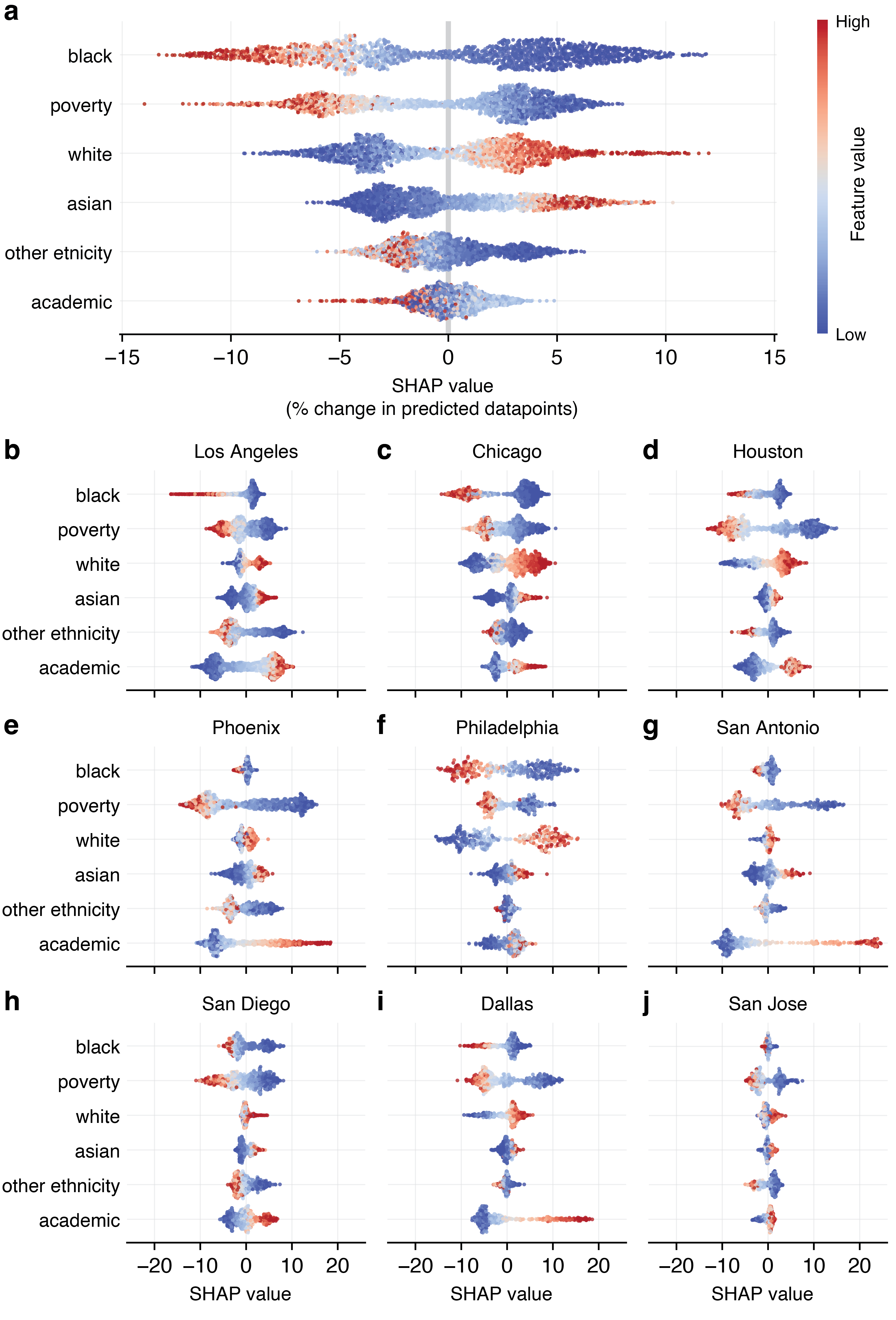}}
\caption{\textbf{Demographic factors which contribute to data production bias}. The figures show SHAP values, which indicate in percent how many data points more (or less) a tract with specific demographic features is predicted to produce relative to the median tract. \textbf{a}, New York City. Variables are arranged in descending order of feature importance. For visualization purposes the plot only shows SHAP values for the most informative variables, see SI Fig. S5-S9 for all variables. The label `academic' denotes the fraction of individuals which have a Bachelor's degree or higher, and the label `other' groups together ethnicities other than Black or African American, White, and Asian into one group. \textbf{b-j}, SHAP values for Los Angeles, Chicago, Houston, Phoenix, Philadelphia, San Antonio, San Diego, Dallas, and San Jose. Note, these plots have slightly enlarged axes compared to panel a. For consistency, variables are ordered according to the arrangement for New York City.}
\label{fig:fig2}
\end{figure}

To uncover how demographic factors are tied with data production we build machine learning regression models to predict how many data points we can expect to observe for a census tract based on its demographic composition.
As features we use demographic variables for ethnicity, poverty, education level, gender, and age.
We approach this from two perspectives, first we build one model per city to understand intricacies within urban areas; later we try to build one general model by merging data across cities.
We use Random Forest models~\cite{breiman2001random} as they are resilient to overfitting and can learn non-linear relationships between variables.
To further avoid overfitting and accurately estimate the performance of the models we use a nested cross-validation scheme (for more details, see Methods).
Building one predictive model per city, we find that demographic factors respectively explain $R^{2}=0.52$ of the variation in data production for New York City, $0.50$ for Los Angeles, $0.57$ for Chicago, $0.53$ for Houston, $0.60$ for Phoenix, $0.45$ for Philadelphia, $0.65$ for San Antonio, $0.45$ for San Diego, $0.56$ for Dallas, and $0.37$ for San Jose (cities are sorted according to population size).
Our goal is not to get a perfect prediction, rather it is to explain which variables contribute to differences in data production.
Fig.~\ref{fig:fig2} shows the SHAP values, which measure how much each variable contributes to the model's prediction~\cite{lundberg2017unified}.
For visualization purposes Fig.~\ref{fig:fig2} only shows SHAP values for the most informative variables (see SI Sec. S3 for SHAP plots including all variables, incl. age, sex, etc.).
Further, to make the figure more informative, SHAP values are rescaled relative to the median census tract (in terms of data production) for each city.
As such, rather than absolute numbers the figure shows the effects demographic factors have in percentages relative to the median tract. 
For New York City (Fig.~\ref{fig:fig2}a) we find that ethnicity, poverty rate, and education levels are the demographic factors which most heavily influence data production.
For example, census tracts with high levels of inhabitants who self-identify as Black or African American have negative SHAP values (dark red dots), ranging from $-5$ to approx. $-13$.
As SHAP values are relative, it means the tract with the highest number of Black or African American individuals in New York City has 13 percentage points less data points relative to the median tract in the city. 
Similarly, tracts with low rates of Black or African American inhabitants have up to 12 percentage points more (dark blue dots).
Put differently, people living in tracts with high percentages of Black or African American individuals tend to produce less mobility data.
This, in turn, reduces their representativity in mobility datasets.

Poverty has a similar effect.
Poorer tracts produce less data (with reductions up to 13\%, Fig.~\ref{fig:fig2}a), relative to the median tract, while wealthy tracts produce up to 8\% more data.
Factors such as the percentage of other ethnicities, and educational levels (the label academic denotes the fraction of individuals which have a Bachelor's degree or higher) are also strongly related to data production levels---albeit lo a lower degree.
We find high levels of individuals who self-identify as White or Asian to be associated with positive SHAP values producing more data relative to the median tract.
However, high levels of `other' ethnicities (consisting of groups other than Black or African American, White, and Asian) and high values of academic (Bachelor's degrees or higher) have predominantly negative SHAP values.
The latter is surprising, as college educated individuals are more likely to own smartphones~\cite{sheet2024pew}, which is necessary to be included in GPS mobility datasets. 
Nonetheless, our results show that tracts with more educated individuals produce up to 7\% less data---demonstrating the complex nuances behind data production bias.

The above results are for New York City, but data from other major US cities (Fig. 2b-j) shows similar robust effects of high levels of poverty and high fractions of Black or African American inhabitants on the number of data points---both are general indicators of low data production. 
However, the strength of these effects varies drastically across cities (see also plot of feature importances SI Sec. S3).
While high levels of poverty reduce data points up to 14.5\% in Phoenix, in San Jose they reduce it only up to 4.9\%.
Similarly, for census tracts with high levels of Black or African American inhabitants effects range from 16.4\% less data points for Los Angeles to 1.4\% less points for San Jose.
High levels of academic degrees, interestingly, have an opposite effect on data production than what we observed for New York. 
We find high percentages of academics living in a tract to have a positive effect on data production with an increase in data production up to 1.6\% for San Jose, 6.2\% for Philadelphia, 6.8\% for San Diego, 8.4\% for Chicago, 9.2\% for Houston, 10.3\% for Los Angeles, 18.5\% for Phoenix, 18.6\% for Dallas, and 24.6\% for San Antonio, relative to the median tract in each city.  
Looking across cities we find that the strongest indicators of data bias differ; for New York the combination of poverty and fraction of Black or African American inhabitants are the most informative features, while for San Antonio it is poverty and academic degree. 
Taken together, these findings show a strong impact of wealth, ethnicity and education on people's data contribution and thus their representation in mobility datasets for all ten major US cities.
However, the considerable variability across cities indicates that the relationship between demographic factors and data bias is not consistent.
Something similar has previously been observed between mobility patterns and socio-economic status, where wealthier groups were found to travel short distances in one city, but longer distances in a different city~\cite{xu2018human}.
Urban factors, like city design, population segregation, etc. seem to have an impact on behavior and data bias.

\subsubsection*{Generalizability of demographic features for estimating data bias}
Our observation of varying relationships between demographic factors and data bias raises the question: is it even possible to build one general model to account for data production bias?
We ask this question as mobility datasets are frequently investigated for biases on national level prior to being used to infer, or model behavior, on finer geospatial scales, such as district, census tract, and city level~\cite{safegraph-bias,chang2021mobility,nilforoshan2023human}.
We therefore investigate how well a model fitted on one, or more, cities performs on an unseen city's demographics. 
We split this investigation into two parts. 
The first part looks at generalizability on a city-to-city basis, i.e. we train a model using data from city A and evaluate on city B.
The second part focuses on training models using data aggregated from 9 cities (out of 10) and testing on the remanding city.
This approach might be more robust and generalizable as it uses a more diverse dataset.

\begin{figure}[!htbp]
\centerline{\includegraphics[width=0.7\linewidth]{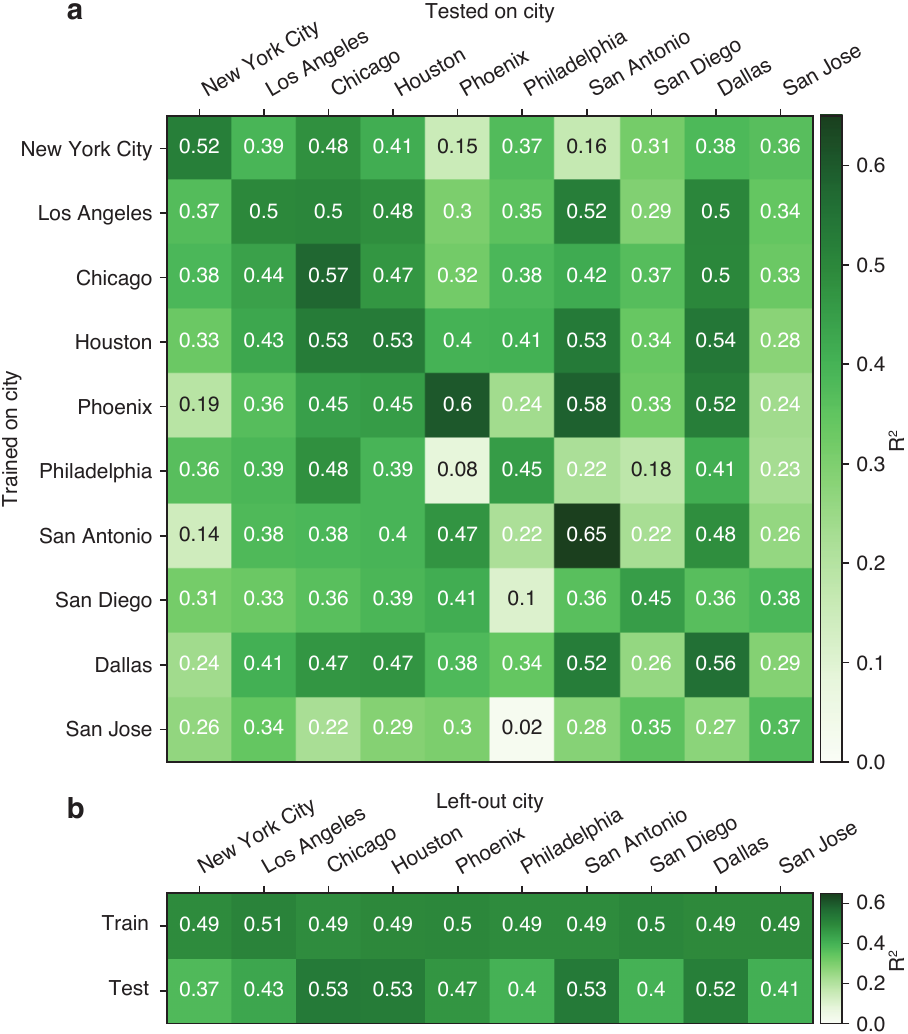}}
\caption{\textbf{Generalization of models for estimating data bias across US cities.} Matrices of model R$^2$-values measuring model performance in predicting median data production of census tracts for unseen cities. \textbf{a}, City-to-city basis of models. Here we train a model on data from one city and test its performance for all other cities. Values in the diagonal (e.g. model trained and evaluated on the same city) are estimated using cross validation. \textbf{b}, Model performance for leave-one-out city models. Here, \textit{train} denotes the achieved performance (cross-validated) on the training set (excluding the city), and \textit{test} indicates the performance on data from the left-out-city.}
\label{fig:fig3}
\end{figure}

Fig.~\ref{fig:fig3}a shows the results for the first approach (train on city A and test on city B).
Models perform generally best for the cities they were originally trained on, however, some city combinations show relatively high R$^2$ values. 
For example, a model trained using Los Angeles data performs better for San Antonio than on previously unseen Los Angeles data.
Nonetheless, we see very little structure in Fig.~\ref{fig:fig3}a, and low values of reciprocity; indicating that if a model trained on city A performs well on data from city B, it does not necessarily mean that a model trained on city B data will perform well on city A. 
Overall, it is difficult to map models trained on one city directly to other cities with good results.
This further underpins our observations that the relationship between demographic factors and data bias differs across cities.
Put differently, cities are highly unique and produce their own specific forms of biases. 

Our second approach investigates whether aggregating data from 9 cities and training a pooled model can improve predictions---we call this the `leave-one-out' city analysis.
As this approach pools data across different cities, covering different aspects of the relationships between demographic factors and data bias, it should make the training data more diverse, and potentially the models more generalizable and robust.
Fig.~\ref{fig:fig3}b shows the results for the leave-one-out city analysis.
Overall, models perform well on the test data from the same city producing R$^2$-values around $0.5$ (top row, Fig.~\ref{fig:fig3}b), however, when evaluated on left-out-cities (bottom row, Fig.~\ref{fig:fig3}b) they produce mixed results.
For half the cities (New York, Los Angeles, Philadelphia, San Diego, and San Jose) the models perform considerably worse on the left-out-city, while, for the other half (Chicago, Houston, Phoenix, San Antonio, and Dallas) model performance is equal to, or better, than during training. 
Comparing left-out-city models (bottom row Fig.~\ref{fig:fig3}b) to individual city models (diagonal values in Fig.~\ref{fig:fig3}a) the leave-one-out city models generally perform worse, except for Houston where the left-out-city model achieves a similar R$^2$ score, and San Jose where it slightly outperforms the individual city model (with a score of R$_{\text{left-out-city}}^2= 0.41$ compared to R$_{\text{San Jose}}^2 = 0.37$).
Comparing left-out-city models (bottom row Fig.~\ref{fig:fig3}b) to individual models trained on other cities (non-diagonal values Fig.~\ref{fig:fig3}a) we find that pooling data across cities does not result in great performance improvements. 
For instance, for New York using data from just Chicago performs better than pooling data across cites.
We observe this for a large majority of cities, e.g. for Los Angeles data from Chicago is better, for Chicago data from Houston is as good, etc. 
Nonetheless, for Houston, San Diego, and San Jose, the pooled data does results in marginal improvements between $0.03$ and $0.05$ percentage point increases in R$^2$.
As such, the benefit of pooling data across cities does not lie in performance gains, rather it lies in eliminating the drawback of having to choose which city to use data from.
This showcases the difficulty in building general models to quantify, and potentially correct, data bias in human mobility data.

\section*{Discussion}
Our increased capacities to collect and analyze large-scale mobility datasets holds great promise in the fight against infectious diseases~\cite{wesolowski2016connecting}, to inform public health actions~\cite{oliver2020mobile}, for humanitarian work and international development~\cite{sekara2024opportunities}, for designing urban environments and transportation networks~\cite{alessandretti2023multimodal}, and for getting a deeper understanding of human actions and behaviors~\cite{schlapfer2021universal,coutrot2022entropy,pappalardo2023future}.
However, for mobility data to truly live up to its transformative potential, and for its benefits to be equally enjoyed by everyone, we need to ensure datasets are representative, equitable, and fair.
In fact, a recent study of GPS data from the US~\cite{li2024understanding} found that groups such as Hispanic populations, low-income households, and individuals with low levels of education are underrepresented in mobility datasets, and identified minor sampling issues for gender, age, and moderate-income.
As we argue here, understanding representativeness is only one aspect of the issue.
Representativeness does not mean we should only strive for individuals to be `represented` in data.
As our results show, how people are represented, or misrepresented, via the quantity of produced data is of equal importance.

We believe our work constitutes a first step towards understanding data generation bias in GPS human mobility.
In our study we focus on a simple metric; how many data points individuals contribute to mobility datasets, and use that as a direct proxy of representation.
Studying mobility for the 10 most populous US cities, we show that data points are unequally produced by individuals, with 60\% of data being produced by only 20\% of people.
To put this into a broader context, the distribution of data points is as unequal as wealth, and for some cities even more extreme.
Similar phenomena have been observed for CDR data~\cite{schlosser2021biases}, albeit not as drastic as we find.
To uncover the underlying causes of data representation we build machine learning models to unpack which demographic factors are related to data production bias.
We find that poverty, ethnicity, and education levels have the largest effect. 
However, the respective impacts of these demographics vary substantially across studied cities.
In turn, this affects the generalizability of the models, and we find no model to perform well on unseen city data.
Pooling data across multiple cities and increases the diversity of training data, but does not alleviate the problem of model generalizability. 
Our approach to model bias is intimately tied with the goals of developing general techniques to reduce bias, or `debias' data.
As our results show, it is difficult to build robust models which can quantify bias geospatially across multiple urban locations. 
Therefore we hypothesize that developing universal debiasing models is unfeasible.

One limiting factor in developing models to predict data production lies in the available demographic data. 
As mobility datsets lack any information on demographic characteristics, we are forced to reduce the resolution of our data from individual to census tract level.
While we do identify relationships between data production and demographic factors, it is difficult to pinpoint them with great accuracy as the resolution of the census data puts a natural limit on this.
Future work can focus on improving this, through either acquiring better data, or using tools which infer demographic characteristics from individual mobility traces. Although we argue against the second approach as it raises serious privacy concerns.

Bias in data production will directly impact the resulting mobility network, and any downstream uses of it, potentially introducing bias in the algorithms which use it.
As biased data has been identified as a key culprit in biased AI systems~\cite{mehrabi2021survey}, it is critical we understand how to address bias in mobility data.
Our approach here is a first step towards understanding this issue.
Due to the nature of data production bias, where it is not only a question of whether specific groups or individuals are included in the dataset, but how they are represented, it is not feasible to use techniques such as post-stratification to correct or remedy this bias.
Future work should investigate how to correct the effects of data production bias on: i) collective scale, such as mobility networks, which are constructed from GPS observations, and ii) individual level.
A complementary approach to address bias can be to look at the data collection process.
Many GPS collected datasets are collected through software development kits which are used by a multitude of apps.
Having information about which apps individual GPS data points stem from (for instance, app type, cost and intended target group) could provide a more complete picture of bias.
As culture has a large influence on which apps people use, and how they use them~\cite{peltonen2018hidden}, having information about this could provide new insights. 

Lastly, we study data production bias in GPS mobility data accessed from a single data provider, however, we believe these biases are not unique to this particular dataset.
They are present across all digital datasets, and we encourage future research to focus on estimating their effects and magnitudes across different human mobility domains.
We urge all practitioners to take these issues seriously. As our results show they can have profound effects, and we recommend everyone who uses mobility data to first do a thorough bias-analysis prior to using the data for other purposes. 

\section*{Methods and Materials}

\textbf{Mobility data.} We use GPS mobility data collected from smartphones by a mobility data provider who requested to be anonymous. They deliver anonymized records of GPS locations from users who have opted-in to provide access to their location data, while remaining compliant under the General Data Protection Regulation and the California Consumer Privacy Act. Our focus is on data generated by individuals living in the ten most populous cities in the United States (NYC, LA, Chicago, Houston, Phoenix, Philadelphia, San Antonio, San Diego, Dallas, San Jose). For each city, the total region of interest is defined as the county in which the city is located (e.g. for Chicago this is Cook County). An exception is NYC, which spans five counties, which are all included in the NYC region of interest (New York County, Kings County, Bronx County, Richmond County, and Queens County). As period of interest, we look at a relatively regular month during which mobility patterns were not affected by COVID travel restrictions: April 2019. \\

\noindent\textbf{Demographic data.} Demographic data from the United States Census Bureau is collected per region of interest at the census tract level from three 2019 American Community Survey (ACS) 5-year estimates subject tables~\cite{acs2019}: Poverty Status in the Past 12 Months (S1701), Age and Sex (S0101) and Educational Attainment (S1501). The last table additionally contains ethnicity information. Census data is given as proportions of the tract population per category. For educational attainment and ethnicity proportions, only population data from people older than 25 is considered, relative to the tract's total 25 plus population.

\noindent\textbf{Data preprocessing.} Initially we extract data records of all active users in the US during the period of interest. We subsequently select users per region of interest (the 10 most populous cities in the US) by filtering them on their inferred home locations, which are identified on the block group level as the user's most frequent location during night. These are mapped to census tracts, such that we for each individual have their `home census tract'. User specific information can then be combined with the census tract-level demographic data on poverty rates, sex, age, educational attainment and ethnicity.
To reduce potential effects of noise we additionally only focus on individual who have produced between $30<$ and $<100,000$ data points during the month. The lower threshold requires individuals to produce at least one data point per day to be included in the study, while the upper one filters away individuals who produce excessive amounts (continually producing more than 2.5 data points per minute). 
In addition, we disregard users if their home census tracts has no associated demographic data in the ACS data, and if tract populations are below 500 inhabitants (to remove noise). 
On average, this removes 2.25\% of individuals (see SI Table S1 for detailed numbers), leaving us with data for approx. 2 million users living across 10.000 census tracts across the 10 cities.

\noindent\textbf{Mobility Network formation.} Nodes represent visited locations at the block group level. Their weight represents aggregated unique user visits. The directional edges represent the aggregated trips between locations, with their weight representing the total trips in that direction. Separate networks were created for the most to the least data contributing user groups; dividing the total user base in 20 equal-user-sized groups, according to approximate similar data contribution withing the period of interest.

\noindent\textbf{Modeling data production.} To study the relative importance of demographic factors we build machine learning models to predict the median data contribution of census tracts.
For each tract we quantify the median data production levels from all individuals having a home location in that specific tract. For the modeling framework we use Random Forest models. A nested K-fold cross validation scheme is used, with three inner folds for hyper parameter tuning and ten outer folds for fitting the model. The coefficient of determination (R$^2$) serves as the training metric. For the leave-one-out exercises, all US city data is used for fitting the model, excluding one city of interest. After training and evaluating each model, it's generalizability can be tested by looking at model performance on unseen city data. We look both at the performance of each city model to every other unseen city data and performance of each leave-one-out model to the left-out city data. For measuring model performance, we use the ordinary least squares coefficient of determination, where model prediction results are evaluated after a possible linear transformation of the outcomes. 

\bibliography{mybib}
%Articles are restricted to 50 references, Letters to 30. No compound references -- only one source per reference.
%When cited in the text, reference numbers are superscript, not in brackets unless they are likely to be confused with a superscript number.
%http://www.nature.com/nature/authors/gta/#a5.4

%% Here is the endmatter stuff: Supplementary Info, etc.
%% Use \item's to separate, default label is "Acknowledgements"

\begin{addendum}
 \item [Acknowledgments] VS and KN are supported by DIREC Denmark (Explore Project, P25). 
 \item[Competing Interests] The authors declare that they have no competing financial interests.
 \item[Correspondence] Correspondence and requests for materials should be addressed to vsek@itu.dk.
\end{addendum}

%"Please note, if you are submitting Supplementary Information with your manuscript, we and our referees prefer it to be submitted as one merged document."
% ---http://mts-nature.nature.com/cgi-bin/main.plex?form_type=display_auth_instructions
%\setcounter{mypostfigure}{0}

\end{document}